*Giant Grüneisen parameter in a strain-tuned superconducting quantum paraelectric: A consequence of the vanishing ferroelectric phonon energy*

J. Franklin[1], B. Xu[1], D. Davino[1], A. V. Balatsky[1,2], U. Aschauer[3], I. Sochnikov[1,4]

[1]*Physics Department, University of Connecticut, Storrs, CT USA, 06269*
[2] *Nordita, KTH Royal Institute of Technology and Stockholm University, Roslagstullsbacken 23, SE-106 91 Stockholm, Sweden*
[3]*Department of Chemistry and Biochemistry, University of Bern, Bern Switzerland*
[4]*Institue of Material Science, University of Connecticut, Storrs, CT USA, 06269*

**Superconductivity and ferroelectricity are typically incompatible because the former needs free carriers, but the latter is usually suppressed by free carriers, unless their concentration is low. In the case of strontium titanate with low carrier concentration, unconventional superconductivity and ferroelectricity were shown to be correlated. Here, we report theoretically and experimentally evaluated Grüneisen parameters whose divergence under tensile stress indicates that the dominant phonon mode that enhances the superconducting order is the ferroelectric transverse soft-mode. This finding rules out all other phonon modes as the main contributors to the enhanced superconductivity in strained strontium titanate. This methodology shown here can be applied to many other quantum materials.**

Lightly doped strontium titanate ($SrTiO_3$) has one of the lowest carrier densities among low carrier-densities superconductors. Its Fermi energy [1,2] is lower than its Debye energy [3,4]. It shows paraelectric-like or ferroelectric-like properties even in the metallic state [5]. Superconducting $SrTiO_3$ cannot be described by a conventional theory of superconductivity and was suggested to be quantum critical [6]. Despite increased research activity, anti-adiabatic quantum-critical low-carrier-concentration superconducting $SrTiO_3$ is still an unresolved fundamental mystery in the field of quantum materials and unconventional superconductors [7–23,1,24,25]. Furthermore, due to its multifaceted nature, $SrTiO_3$ can be useful in designing remarkable new functionalities [26].

On the theory side, the debate focuses on the superconducting pairing mechanism. Specifically, there is much interest in what phonons provide the superconducting pairing and how [27–35]. Some recent experiments on approaching the quantum paraelectric to ferroelectric phase transition show correlations of the ferroelectric phase with superconductivity [19,20,36–39]. However, the observations so far are rather qualitative, and have only shown enhanced critical temperatures and thus a connection between the two quantum phases. These findings are difficult to be interpreted quantitatively on a microscopic mechanistic level. The current work provides a clear quantitative link between the macroscopic and microscopic views of the interacting ferroelectric and superconducting phases in this enigmatic material.

The Grüneisen parameter describes the effect of a changing crystal lattice geometry on its vibrational properties. The Grüneisen parameter has been successfully used in the past to unravel details about quantum phase transitions [40–43]. In SrTiO$_3$, to understand the relationship between strain and the superconducting T$_c$ near the quantum paraelectric phase transition on a quantitative level, we investigate the relevant experimentally measured uniaxial strain Grüneisen parameter defined as $\widetilde{\gamma^c}(\varepsilon) = \frac{dT_c}{T_c(\varepsilon)d\varepsilon_c}$. $\varepsilon_c$ is the measured strain along the c-axis and $T_c$ is the superconducting transition temperature [34,36,38]. The two most striking breakthroughs we report here are: (a) the Grüneisen parameter in SrTiO$_3$ reaches gigantic values on the order of several thousands, larger than previously reported in many conventional and unconventional materials [4–9], and (b) these values agree with our theoretical calculations within the inspected strain ranges. These findings provide a clear consistent evidence that the response of the superconducting phase to strain is predominantly due to the soft, so-called transverse optic (TO), ferroelectric phonon mode with displacements along the tensile strained c-axis direction. Furthermore, these findings also show that no other phonons, including the longitudinal ferroelectric branches, contribute noticeably to the anomaly in the response.

Here, the mode Grüneisen parameters for the phononic response to uniaxial strain are defined as $\gamma_{n,ij,\boldsymbol{q}}(\varepsilon) = -\frac{1}{\omega_{n,\boldsymbol{q}}}\frac{\partial \omega_{n,\boldsymbol{q}}}{\partial \varepsilon_{ij}}$, where $\omega_{n,\boldsymbol{q}}$ is the frequency of a phonon mode $n$, $i$ and $j$ are the strain tensor indices and $\boldsymbol{q}$ is the phonon wavevector [51]. In experiments under uniaxial *stress* the modes' responses are superimposed as oftentimes the experimental strain is not in its irreducible form. Thus, a superposition of the uniaxial *strain* mode Grüneisen parameters should be used. For our experiments a uniaxial stress along the c-axis is applied and the strain in the same direction is monitored [1–3]. Assuming weakly coupled phonon modes and a positive Poisson ratio under given strain-stress conditions, we can approximate [53] the c-axis Grüneisen parameter as $\gamma^c(\varepsilon) = \sum_{n,\boldsymbol{q}}\left(\gamma_{n,\boldsymbol{q}}^c(\varepsilon)\right) \approx \sum_{n,\boldsymbol{q}}\left(-\frac{(1-2v)}{\omega_{n,\boldsymbol{q}}}\frac{\partial \omega_{n,\boldsymbol{q}}}{\partial \varepsilon_c}\right)$, where $v \approx 0.28$ is the low temperature Poisson's ratio for SrTiO$_3$ [46], and the sum is over all phonon modes $n$ and wave vectors $\boldsymbol{q}$. This expression assumes isotropy even though strontium titanite is not isotropic. This assumption is satisfactory for emphasizing the main point about the origin of the anomalous response from the ferroelectric soft-mode. For a more complex and rigorous theoretical treatment of anisotropic materials see M. Mito *et al.* [45]. We will make a further simplification in the following discussion by inspecting the Grüneisen parameters at the Γ point only, i. e. $q = 0$. This is motivated by the fact that in our computational supercell the cubic zone-boundary modes are folded back to the Γ point, and also by the fact that no anomalous response to strain was detected at intermediate wave-vectors in our calculations. The two Grüneisen parameters defined above, $\widetilde{\gamma^c}$ and $\gamma^c$, are expected to be approximately equivalent near the superconducting phase transition if it is assumed that the electronic behavior is dominated by a single energy scale [40] of the superconducting pairing, $T_c$, and that the electronic and phononic properties are linked through electron-phonon coupling [35,55,56]. This equivalence is what allows for direct conclusions about the microscopic phonon behavior based on the thermodynamic bulk $T_c$ measurements under strain.

DFT calculations, the results of which are shown in Figure 1, were performed within the VASP code [57–60] using the PBEsol exchange correlation functional [61] and projector-augmented wave (PAW) potentials [62,63] with Sr(4s, 4p, 5s), Ti(3p, 3d, 4s) and O(2s, 2p) valence electrons. Wavefunctions were expanded in planewaves up to a kinetic-energy cutoff of 550 eV. We relaxed the lattice parameters and internal coordinates until forces converged below $10^{-5}$ eV/Å and stress converged below $5 \cdot 10^{-7}$ eV/Å$^3$. All calculations were performed for 40-atom 2x2x2 supercells of the cubic unit cell that contained the tetragonal distortion corresponding to an $a^0 a^0 c^-$ rotation of the octahedra in Glazer notation [64]. Reciprocal space was sampled using 4x4x4 Monkhorst-Pack [65] mesh for this supercell. Phonon frequencies were computed at the Γ-point within the frozen phonon approach, which was implemented in the Phonopy code [65]. Mode Grüneisen parameters were computed as $\gamma_n^c(V) = -\frac{V}{\omega_n}\left(\frac{\partial \omega_n}{\partial V}\right)_c = -\frac{1}{\omega_n}\frac{\partial \omega_n}{\partial \varepsilon_c} \equiv \frac{\gamma_{n,q=0}^c(\varepsilon)}{(1-2\upsilon)}$, where $V$ is the volume. The derivative was evaluated by central finite differences, connecting modes at adjacent volumes via the similarity of their eigenvectors. The denominator with the Poisson's ratio, $\upsilon$, allows the translation from the Grüneisen parameter calculated from volume $\gamma_n^c(V)$ to strain $\gamma_n^c(\varepsilon)$. The longer side of the tetragonal unit cell of strontium titanate is defined as the *c* axis (Figure 1). The volume was modulated by applying strain along the *c* axis, and then relaxing the cell shape, volume and all internal coordinates while keeping *c* fixed.

Within the investigated range of *c* lattice parameters only the ferroelectric mode along *c* changes from unstable (negative) at large *c* to stable at small *c*, see Figure 1 (upper panel). All other modes show a much smaller dependence on strain along *c*. Notably, the doubly degenerate in-plane ferroelectric mode remains unstable at all *c*. These observed instabilities are in agreement with the quantum paraelectric nature of SrTiO$_3$ that manifests as unstable phonon modes in our 0K DFT calculations without zero-point energy corrections [66,67]. Consequently, the Grüneisen parameter (Figure 1, lower panel) of the ferroelectric mode along *c* is much larger, by two to three orders of magnitude, than the one for all other modes and shows the divergence expected from its definition at the critical *c* lattice parameter [40]. We want to note that the modes shown in this figure also include zone boundary modes such as octahedral rotations because these modes are folded back to the Γ-point within our supercell. Because of the relevant magnitude of the Grüneisen parameters, these findings rule out other than the ferroelectric transverse phonon modes as main contributors to enhanced superconductivity.

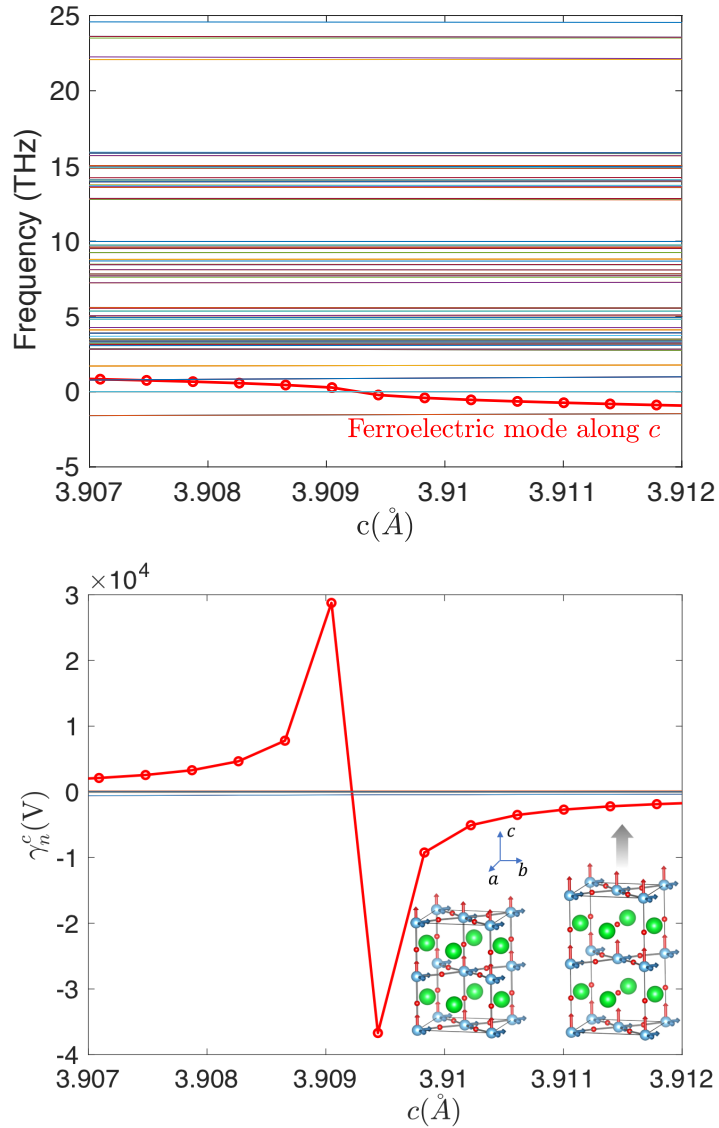

*Figure 1. Theoretical phonon frequencies (top) and Grüneisen parameters calculated (bottom) for phonon modes, $\gamma_n^c(V)$, depending on the c lattice parameter. The ferroelectric mode with vibrations along the c-axis is highlighted in red, while all other modes are shown in grey. Only the transverse ferroelectric mode along the c-axis shows large values of the Grüneisen parameter (bottom).*

*The inset shows an illustration for unstrained strontium titanate unit cell (left) and the strained one (right). Red, green and blue spheres are oxygen, strontium and titanium ions respectively. The blue and red errors indicate the relative amplitudes of perpendicular and parallel to c-axis branches of the transverse ferroelectric mode, respectively. The parallel to the c-axis branch of the transverse ferroelectric mode acquires larger amplitude due to the phonon softening under the tensile deformation indicated by the gray arrow.*

Experiments were performed on single crystals of Nb-doped strontium titanate, $SrTi_{0.996}Nb_{0.004}O_3$, in a dilution refrigerator setup with a custom-built strain-stress cell [68] and a polarizing optical microscope [38]. Details of the basic characteristics of the samples are provided elsewhere [38]. Ultra low excitation currents and an ultra-low noise amplifier and resistance bridge were employed [39,52] to determine the critical temperature, $T_c$. Stress was applied parallel to the long side of a 0.3 x 2 x 10 mm³ single crystal sample to define *c* axis. The sample's resultant strain along the stress direction was precisely measured using an attached resistive strain gauge.

We measured the resistive signature of the superconducting transition and defined $T_c$ at different normal resistance, $R_n$, thresholds. Typical critical temperature data are shown in Figure 2 (upper panel). Overall $T_c$ increased by ~30% before the sample fractured. What is remarkable is that the change happens over a very small range of induced strain. In other words, the derivative of $T_c$ with respect to strain, which yields the Grüneisen parameter $\widetilde{\gamma^c}(V) \equiv \widetilde{\gamma^c}(\varepsilon)/(1-2\upsilon)$, is anomalously large (see Figure 2, lower panel).

Moreover, when sufficiently large strains were achieved, we observed a non-linear upturn in $T_c$ and the Grüneisen parameter. Such divergence would happen if a system is pushed towards a (quantum) phase transition as one expects in $SrTiO_3$ [4,69–71]. Comparing this experimental finding (Figure 2) with the theory (Figure 1) shows remarkably similar large divergent values and supports the microscopic ferroelectric soft-mode picture underlying the superconducting mechanism in $SrTiO_3$.

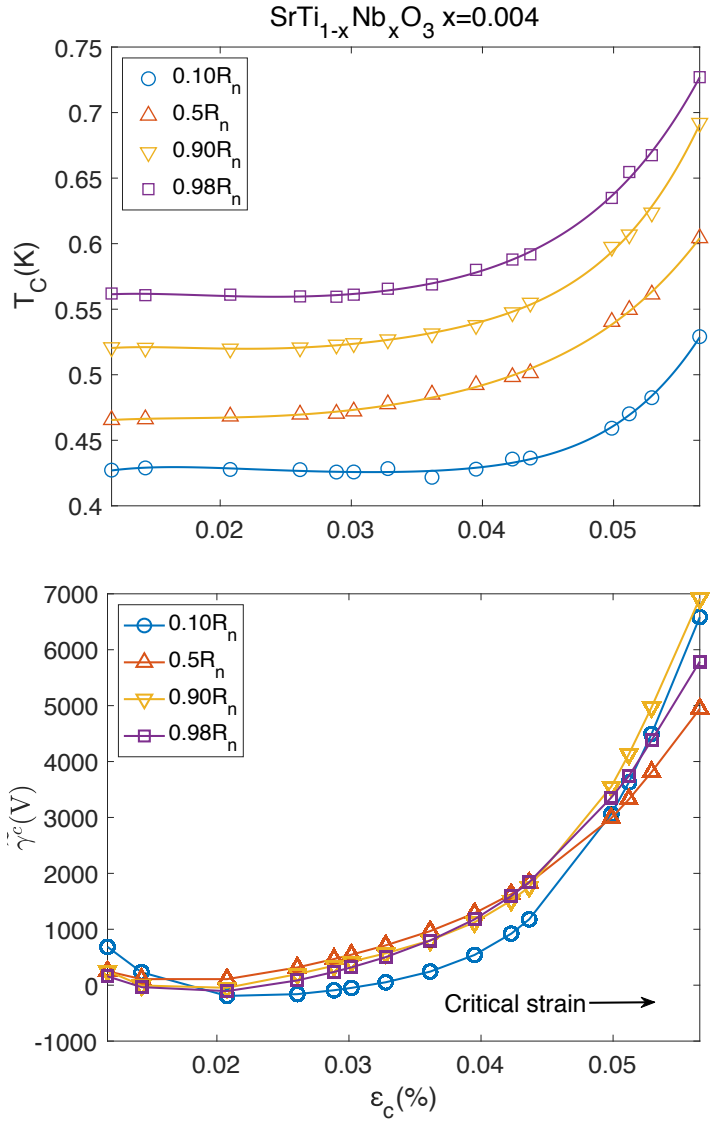

*Figure 2. Experimental anomalous Grüneisen parameter $\widetilde{\gamma^c}(V)$ determined from numerical derivative of superconducting $T_C$ for Nb doped SrTiO₃. The crystals were strained in the c-axis direction using uniaxial tension. The derivative is based on 6-th degree polynomial smoothing fits (solid lines in the top panel). The Grüneisen parameter (bottom panel) shows anomalous growth reaching several thousand at maximal attained strains, which is higher than in many conventional and unconventional superconductors (lines are numerical derivatives of the fits to the experimental data in the upper panel, symbols are a guide for the eye to show at which strains $T_c$ was actually measured).*

This is a striking finding. In conventional superconductors and even unconventional superconductors [46,47], the Grüneisen parameter based on $T_c$ is typically a few tens (unitless) [46–50]. It can be larger, many hundreds, only when another phase transition occurs, like in topologically non-trivial changes in the electronic bands [45].

In summary, the divergence of the calculated Grüneisen parameter corresponds to the divergence of the experimental superconducting critical temperature. This deserves special attention because it indicates a non-trivial pairing scenario associated with sub-picometer structural displacements, that push the system toward the paraelectric-ferroelectric structural phase transition [43]. Relating the anomalous response to strain to the anomalous mode Grüneisen parameter essentially states that the ferroelectric soft mode must be the key element in a theory that can correctly describe the pairing. This is true whether that theory is a direct [7,8], multi-photon [35], screening [72] or any other scenario. According to our findings, any proposed mechanisms would have to have the pairing energy depend strongly on the ferroelectric mode even if they include some coupling to other phonon modes. The methods presented here can be applied to numerous other quantum materials such as high temperature and unconventional superconductors [73–76], quantum magnets [77–79], and topological matter [80–82]. Beyond this report, interesting future experiments with induced strains in quantum materials may include testing directly the mode Grüneisen parameters in scattering experiments [83–85] and determining a mesoscale and nanoscale response using microscopies such as a scanning SQUID [86].

Acknowledgements

We thank J. N. Hancock and V. Juričić for valuable suggestions. IS acknowledges the US State of Connecticut and the US DOD for partial support. UA was supported by the Swiss National Science Foundation Professorship Grants PP00P2_157615 and PP00P2_187185. Calculations were performed on UBELIX (http://www.id.unibe.ch/hpc), the HPC cluster at the University of Bern and at the Swiss Supercomputing Center (CSCS) under project s955. AB was supported by VILLUM FONDEN via the Centre of Excellence for Dirac Materials (Grant No. 11744), the European Research Council under the European Union's Seventh Framework Program Synergy HERO, and the Knut and Alice Wallenberg Foundation KAW 2018.0104.